\def\year{2022}\relax
\documentclass[letterpaper]{article} 
\usepackage{aaai22}  
\usepackage{times}  
\usepackage{helvet}  
\usepackage{courier}  
\usepackage[hyphens]{url}  
\usepackage{graphicx} 
\usepackage{natbib}  
\usepackage{caption} 
\DeclareCaptionStyle{ruled}{labelfont=normalfont,labelsep=colon,strut=off} 
\usepackage{algorithm}
\usepackage{algorithmic}

\usepackage{booktabs}
\usepackage{multirow}
\usepackage{amsmath, amssymb, amsfonts}
\usepackage[toc,page]{appendix}

\usepackage{newfloat}
\usepackage{listings}
\floatstyle{ruled}
\newfloat{listing}{tb}{lst}{}
\floatname{listing}{Listing}
\title{SR-GCL: Session-Based Recommendation with Global Context Enhanced Augmentation in Contrastive Learning}

\author{
    Eunkyu Oh, Taehun Kim, Minsoo Kim, Yunhu Ji, Sushil Khyalia
}
\affiliations{
    Samsung Research, Samsung Electronics\\
    Seoul, Republic of Korea\\
    \{eunkyu1.oh, taehun33.kim, minsoo3.kim, yunhu.ji, sus.hil\}@samsung.com
}

\begin{document}
\maketitle

\begin{abstract}
Session-based recommendations aim to predict the next behavior of users based on ongoing sessions. The previous works have been modeling the session as a variable-length of a sequence of items and learning the representation of both individual items and the aggregated session. Recent research has applied graph neural networks with an attention mechanism to capture complicated item transitions and dependencies by modeling the sessions into graph-structured data. However, they still face fundamental challenges in terms of data and learning methodology such as sparse supervision signals and noisy interactions in sessions, leading to sub-optimal performance. In this paper, we propose SR-GCL, a novel contrastive learning framework for a session-based recommendation. As a crucial component of contrastive learning, we propose two global context enhanced data augmentation methods while maintaining the semantics of the original session. The extensive experiment results on two real-world E-commerce datasets demonstrate the superiority of SR-GCL as compared to other state-of-the-art methods.
\end{abstract}

\begin{figure*}[t]
    \centering
    \includegraphics[width=2\columnwidth,keepaspectratio]{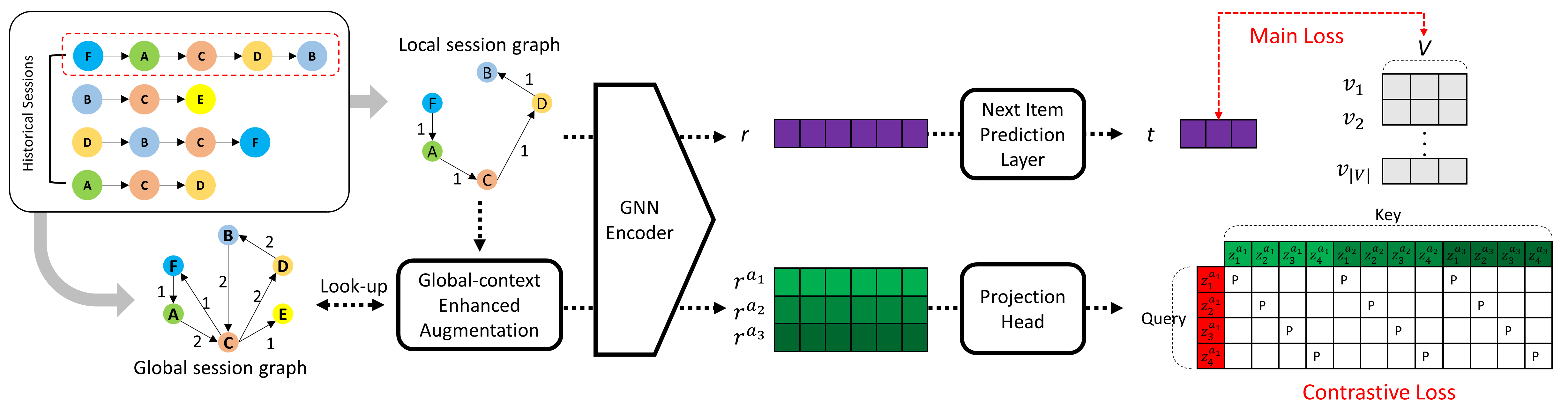}
    \caption{Overall flow of SR-GCL. Augmentations selected from a method pool are applied to a local session graph. Then, the GNN-based encoder, Next Item Prediction Layer, and Projection Head are jointly optimized with two different tasks.}
    \label{fig:archi}
\end{figure*}

\section{Introduction}
A Session-Based Recommendation (SBR) has attracted increasing attention as a new paradigm for Recommender Systems (RecSys) in E-commerce and online streaming services, which aims to predict successive items that a user is likely to interact with given a sequence of previous items consumed in the session. By definition, a session is a slice of user-item interactions separated by arbitrary time, so it is naturally expressed as a variable-length time-series event sequence \cite{wang2019survey}. With the advances of deep learning techniques, Graph Neural Networks (GNNs)-based approaches \cite{wu2019session, pan2020star} show the effectiveness to capture sophisticated transition relationships of items within the given session.

Although existing methods achieve promising results, they still suffer from some limitations in terms of data and learning methodology. Firstly, they mostly follow the supervised learning paradigm \cite{hidasi2015session, wu2019session} where the supervision signal comes from observed user-item interactions. However, as well known problems in RecSys, the observed data are extremely sparse \cite{bayer2017generic, he2016ups} compared to the whole interaction space, and usually follow a power-law distribution \cite{milojevic2010power}.
Due to these data characteristics, high-degree items receive more supervision signals, while it is hard to obtain high-quality representations of low-degree ones.
Secondly, they are vulnerable to noisy interactions that deviate from the user's main intention, especially in the case of a long-range session \cite{wang2019survey}.

To overcome the aforementioned problems, we adopt Contrastive Learning (CL) that has renewed a surge of interest in Computer Vision (CV) and Natural Language Processing (NLP) domains \cite{hjelm2018learning, wu2018unsupervised}. This leverages the fact that the underlying data has a richer structure than the information that sparse labels or rewards can provide. Specifically, CL utilizes data- or task-specific augmentations to inject the desired feature invariance, distills additional supervision signals from the unlabelled data itself, and represents semantically similar objects (i.e., positive pairs) closer to each other, while dissimilar ones (i.e., negative pairs) further away. In this process, data augmentation is an essential component and has been researched for images \cite{perez2017effectiveness} (e.g., color jitter, random flip, etc.) and texts \cite{wei2019eda} (e.g., synonym replacement, random insertion, etc.). In contrast, an augmentation scheme for an SBR has not been sufficiently studied. As an alternative, applying conventional ones into a CL framework for an SBR is also not straightforward because such perturbation can loose the context of the whole session.

Along these lines, we propose SR-GCL, a novel contrastive learning framework for a session-based recommendation. To be specific, each session is firstly constructed to a directed graph as an anchor, then augmented session graphs are generated as positive pairs by choosing arbitrary augmentation methods. We also obtain negative pairs by augmenting other samples. In this process, we propose two data augmentation methods for maintaining the semantics of the original session. For this, we consider the global context by integrating all pairwise item transitions over whole sessions. In this way, the influence of noisy items and high-degree nodes can be reduced, so that accurate neighbor information of nodes can be extracted. Our model is optimized by a next-item prediction objective and multi-positive contrastive learning simultaneously. Extensive experiments on two public datasets demonstrate the effectiveness of our framework.

\section{Related Work}
\subsection{Session-Based Recommendation}
In an SBR, a session is formally represented as a variable-length time-series event sequence. With the outstanding performance of deep learning-based models, RNN-based methods model a series of events in the session as a sequence and take the last hidden state as the context representation \cite{hidasi2015session}. Recently, GNNs have become dominant, which more accurately capture the transition pattern and coherence of items within a session by modeling the session sequences as graph-structured data \cite{wu2019session, pan2020star}. Despite the remarkable success, many existing works following a supervised learning paradigm still suffer from data sparsity and a long-tail problem in user-item interactions \cite{bayer2017generic, clauset2009power}. To address the weak supervision from insufficient labels and to better learn the optimal representation of items and sessions, we adopt a contrastive learning paradigm for an SBR.

\subsection{Contrastive Self-Supervised Learning}
Recently, Self-Supervised Learning (SSL) has shown impressive performance in various fields while complementing the limitations of supervised learning.
Among SSL, CL builds representations by learning to encode what makes two things similar or different via positive or negative samples \cite{he2020momentum, chen2020improved, chen2020simple, chen2020big}.
For sequential recommendations, \cite{xie2020contrastive} proposed Contrastive Learning for Sequential Recommendation (CL4SRec) extracting more meaningful user patterns within a session and three augmentation approaches such as Crop, Mask, and Reorder. However, the augmentation methods have no consideration of session data characteristics as those are merely borrowed from other domains (e.g., image). Unlike the previous methods, our novel augmentation approaches retain the original intention of a given session by taking into account the global context.

\section{Method}
\subsection{Contrastive Learning Framework}
Figure \ref{fig:archi} presents the overview of our contrastive learning framework consisting of four major components: Augmentation, GNN Encoder, Next Item Prediction Layer, and Projection Head. The proposed framework builds representations of items and a session (i.e., a sequence of items) while maximizing agreement between different views on the session via a data augmentation pipeline. The trainable modules are jointly optimized with two different tasks: a next item prediction as the main task on which most of the previous models have focused and contrastive learning as an auxiliary task.

For the next item prediction task, each session $s$ is first constructed as a directed graph and fed to an encoder, which computes the representation $r$ of the session. In our case, $f(\cdot) : \mathbb{R}^{\left\vert s \right\vert \times d} \to \mathbb{R}^{2d}$ denotes the encoder, where $\left\vert s \right\vert$ is the length of $s$ and $d$ is the dimension of an item embedding. In practice, the proposed framework is model-agnostic and can adopt any encoder for session-based recommendations. In this paper, we adopt SGNN-HN, a state-of-the-art GNN-based model \cite{pan2020star} (See Appendix \ref{Appendix_SGNN} for the detail). The representation of a session via the encoder can be expressed as
\begin{equation} \label{eq:r}
    r = f(s)
\end{equation}
Then, Next Item Prediction Layer uses $r$ to obtain a predicted next click $t$. $g(\cdot) : \mathbb{R}^{2d} \to \mathbb{R}^{d}$ denotes the prediction function, which consists of one linear layer as
\begin{equation} \label{eq:t}
    t = g(r) = W_g r + b_g    
\end{equation}
where $W_g \in \mathbb{R}^{d \times 2d}$ is a trainable matrix and $b_g \in \mathbb{R}^{d}$ is a bias. We obtain main loss $\mathcal{L}_{main}$ by measuring the similarities between $t$ and all items $V$.

On the other hand, the process of contrastive learning starts with augmentations. Our framework has no restriction on how many augmentation methods are used per batch, but we assume two randomly selected augmentations for simplicity \cite{he2020momentum, chen2020improved, chen2020simple, chen2020big}. Augmentation module generates two augmented sessions, $s^{a_i}$ and $s^{a_j}$ by each method. Thereafter, we obtain their representations, $r^{a_i}$ and $r^{a_j}$ by passing them through the same encoder with shared parameters as Equation \ref{eq:r}. Projection Head (See Appendix \ref{Appendix_projection_head} for the detail) represented by $h(\cdot) : \mathbb{R}^{2d} \to \mathbb{R}^{d}$ maps the representations into another embedding space, and outputs each $z$ as
\begin{equation} \label{eq:z}
    z = h(r) = W_{h2} \sigma(W_{h1}r + b_{h1}) + b_{h2}
\end{equation}
where $W_{h1} \in \mathbb{R}^{d \times 2d}$, $W_{h2} \in \mathbb{R}^{d \times d}$ are learnable matrices, $b_{h1}, b_{h2} \in \mathbb{R}^d$ are biases, and $\sigma$ is a ReLU. Finally, we calculate contrastive loss $\mathcal{L}_{cl}$ through comparison with all $z$ obtained in the same batch.

\begin{figure}[t]
    \centering
    \includegraphics[width=0.48\textwidth,keepaspectratio]{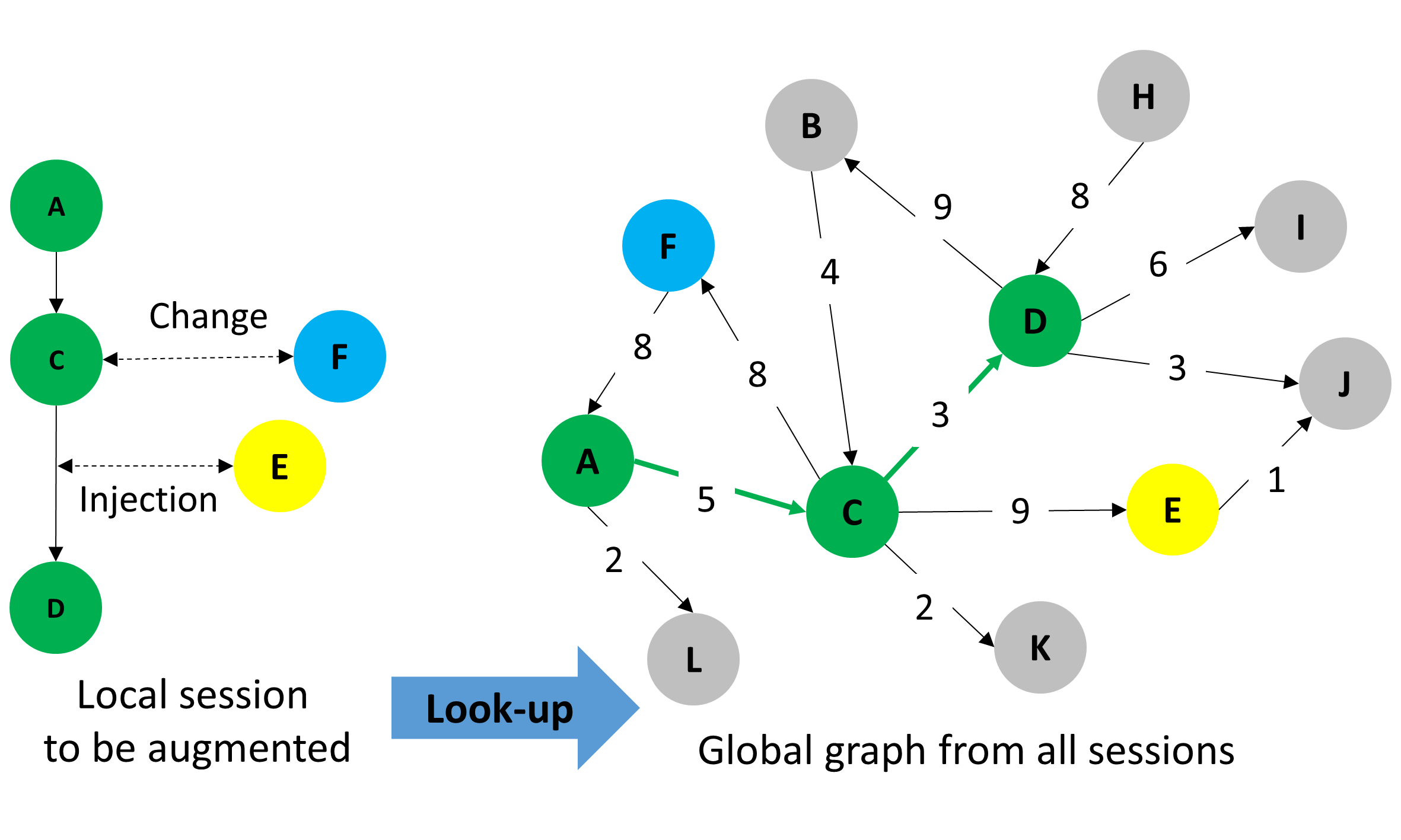}
    \caption{Global context enhanced augmentation.}
    \label{fig:augmentation}
\end{figure}

\subsection{Data Augmentations for a Session}
A successful session augmentation should be able to obtain different views of the same sequence but still, maintain the main characteristics hidden in historical behaviors. Inspired by the principles of augmentations in NLP \cite{wei2019eda, jiao2019tinybert}, we devise two global context enhanced augmentation approaches (i.e., Item Change and Injection) by referring to the global graph having pairwise connections over whole sessions as shown in Figure \ref{fig:augmentation}. Synonyms of a node are sampled by the degree of connections from the global graph. These methods can reflect the global intra-session context, not just the given session.

Our proposed approaches have two main advantages compared with previous methods (See Appendix \ref{Appendix_augmentations} for the details of conventional methods for session data). First, the conventional methods mainly deal with limited information in a current session, whereas our approaches considering the entire session context information can generate more plausible session data. Second, we can obtain many variants of the original sample. If we should augment only within the given session, the maximum number of variants is restricted. These advantages are particularly pronounced for session data with a short length.

For common notations, $\gamma \in \left[ 0, 1 \right]$ means an augmented ratio. $\delta(x)$ is the function that picks one node from the neighbors of a given item $x$. The neighbor nodes are termed synonyms, which have similar interactions with $x$. $\delta(x)$ samples the synonym with the frequency of connections to the power of $k \in \left[ 0, 1 \right]$ to lower the impact of high-degree nodes with respect to a long-tail problem.

\subsubsection{Item Change}
Item Change is to replace a selected item with its synonym. To apply this augmentation to a session $s$, we randomly make a set $\mathcal{I}_{ch} = \left\{ i_1, i_2, ..., i_{N_{ch}} \right\}$ indicating changed indices, where $N_{ch} = \lceil \gamma_{ch} \times \left\vert s \right\vert \rceil$. The items pointed by $\mathcal{I}_{ch}$ are changed to their synonyms. This method $a_{ch}$ is formulated as
\begin{equation} \label{eq:ch}
\begin{aligned}
    s^{a_{ch}} = a_{ch}(s) = \left[ \tilde{x}_1, \tilde{x}_2, ..., \tilde{x}_{ \left\vert s \right\vert } \right], \\
    \mbox{where } \tilde{x}_j = 
    \begin{cases}
    x_j \qquad\qquad \mbox{if } j \notin \mathcal{I}_{ch}\\
    \delta(x_j) \qquad\;\;\; \mbox{if } j \in \mathcal{I}_{ch}
    \end{cases}
\end{aligned}
\end{equation}

\subsubsection{Item Injection}
This augmentation is to insert synonyms into random indices of a sequence. Formally, this inserts an item $N_{in} = \lceil \gamma_{in} \times \left\vert s \right\vert \rceil$ times to the random positions of a session $s$. When $i(s)$ denotes the function that one synonym of a randomly selected item is injected before or after the item, this method $a_{in}$ repeats $i(\cdot)$ $N_{in}$ times (i.e., $i^{N_{in}}(\cdot)$), which can be formulated as
\begin{equation} \label{eq:in}
\begin{aligned}
    s^{a_{in}} = a_{in}(s) = i^{N_{in}}(s), \qquad\quad\\
    \mbox{where } i(s) = \left[ x_1, ..., x_i, \delta(x_{i}), x_{i+1}, ..., x_{\left\vert s \right\vert} \right]
\end{aligned}
\end{equation}

\subsection{Objective Function for Multi-task Learning}
We take multi-task learning to enhance the performance of a session-based recommendation. The main task for a next item prediction and an auxiliary task for contrastive learning are jointly optimized. The total loss $\mathcal{L}_{total}$ is the weighted sum of main loss $\mathcal{L}_{main}$ and contrastive loss $\mathcal{L}_{cl}$ as
\begin{equation}
    \mathcal{L}_{total} = \mathcal{L}_{main} + \lambda\mathcal{L}_{cl} 
\end{equation}
where $\lambda$ is a weight for contrastive loss.

The process of deriving $\mathcal{L}_{main}$ for the prediction task is as follows. A session $s$ is encoded to the representation via Encoder as Equation \ref{eq:r}, and a predicted next click $t$ is obtained via Next Item Prediction Layer as Equation \ref{eq:t}. Then, we compare the predicted click $t$ with all items in $V$ by calculating the similarities. To solve a long-tail problem in a recommendation \cite{abdollahpouri2017controlling}, we apply layer normalization before dot-product as
\begin{equation} \label{eq:dot_product}
\begin{aligned}
    sim(v_1, v_2) = \tilde{v}_1^\top \tilde{v}_2 \quad\\
    \mbox{where } \tilde{v} = LayerNorm(v)
\end{aligned}
\end{equation}
To normalize the similarities, a softmax layer is applied with a temperature parameter $\tau_m$. Finally, we adopt cross-entropy as an optimization objective. These are summarized as 
\begin{equation}
    \mathcal{L}_{main} = -\log\frac{\exp(sim(t, y)/\tau_m)}
    {\sum_{v_i \in V}\exp(sim(t, v_i)/\tau_m)}
\end{equation}
where $y \in V$ is the next clicked item of $s$.

For contrastive loss, we extend a Simple framework for Contrastive Learning of visual Representations (SimCLR) \cite{chen2020simple} that does not require a separate module to manage negative samples. However, the number of negative samples is determined by the number of mini-batch $N$ since the negative samples are generated from other samples in the same batch. If more negative samples are required for enhancing a model, $N$ should also increase. To overcome this dependency, we propose multi-positive contrastive loss inspired by supervised contrastive loss \cite{khosla2020supervised}. This allows us to flexibly set the number of negative samples, which can employ multiple augmentation methods per batch (i.e., more than two). That is, we generalize InfoNCE loss \cite{oord2018representation} based on noise contrastive estimation for the multiple positive pairs.

The procedure to compute multi-positive contrastive loss $\mathcal{L}_{cl}$ is as follows. Augmentation module generates $M \times N$ augmented sessions, where $M$ is the number of selected augmentations. Then, the augmented sessions are encoded to the representations as Equation \ref{eq:r}, and those are projected into another space via Projection Head as Equation \ref{eq:z}. For example, when a session $s_n$ augmented by $a_m$ passes Encoder $f(\cdot)$ and Projection Head $h(\cdot)$, it becomes $z_n^{a_m}$. Therefore, we obtain a set $Z^\prime = \left\{z_1^{a_1}, ..., z_N^{a_1}, ..., z_1^{a_M}, ..., z_N^{a_M} \right\}$. To compute $\mathcal{L}_{cl}$, we define a query set $Z_q^\prime = \left\{ z_1^{a_1}, z_2^{a_1}, ..., z_N^{a_1} \right\}$, which contains $N$ representations of the sessions augmented by the first method $a_1$. For each query, the similarities with $M \times N$ keys $Z_k^\prime = Z^\prime$ are calculated as Figure \ref{fig:archi}. The same distance as Equation \ref{eq:dot_product} is used to measure the similarities of the query and keys. A softmax layer with a temperature parameter $\tau_c$ is applied. Cross-entropy is normalized for the multiple positives. That is, multi-positive contrastive loss for the $q$-th query $z_q \in Z_q^\prime $ is so defined as
\begin{equation} \label{eq:multi_positive}
    \mathcal{L}_{cl} = \frac{-1}{M}\!\!
    \sum_{z_{pos} \in P(z_q)}\!\!\!\!\! \log
    \frac{\exp(sim(z_q, z_{pos})/\tau_{c})}
    {\sum_{z_k \in Z_k^\prime}\exp(sim(z_q, z_k)/\tau_{c}) }
\end{equation}
where $P(z_q) \subseteq Z_k^\prime$ is a set of all positive pairs of the $q$-th query, and $M = \left\vert P(z_q) \right\vert$ is the number of selected methods.

\section{Experiments and Analysis}
\subsection{Experimental Settings}
\subsubsection{Datasets}
\begin{itemize}
\item Yoochoose was released by RecSys Challenge 2015\footnote{https://recsys.acm.org/recsys15/challenge/}, which contains click-streams from an E-commerce website within 6 months.
\item Diginetica was used as a challenge dataset for CIKM Cup 2016\footnote{https://competitions.codalab.org/competitions/11161}. We only adopt the transaction data which is suitable for a session-based recommendation.
\end{itemize}

We follow the preprocessing of previous works \cite{wu2019session, pan2020star} for fairness. We filter out sessions of length 1 and items which occur less than 5 times. Then, we split the sessions for train and test, where the last day of Yoochoose and the last week of Diginetica are used for test. Furthermore, we exclude the items which are not included in the train set. Finally, we split the sessions to several sub-sequences. Specifically, for a session $s = \left[ x_1, x_2, ..., x_{\left\vert s \right\vert} \right]$, we generate sub-sequences and the corresponding labels as $\left( \left[ x_1 \right], x_2 \right), \left( \left[ x_1, x_2 \right], x_3 \right), ..., \left( \left[ x_1, ..., x_{\left\vert s \right\vert - 1} \right], x_{\left\vert s \right\vert} \right)$ for train and test. As Yoochoose is too large, we only utilize the recent 1/64 and 1/4 fractions of the train set, denoted as Yoochoose1/64 and Yoochoose1/4, respectively. Statistics for the three datasets are summarized in Appendix \ref{Appendix_datasets_characteristics}.

\subsubsection{Baselines}
\begin{itemize}
\item \textbf{GRU4Rec} \cite{hidasi2015session} applies GRUs to model the sequential information in a session-based recommendation.
\item \textbf{CSRM} \cite{wang2019collaborative} employs GRUs to model the sequential behavior, adopts an attention mechanism to capture the main purpose, and uses neighbor sessions as auxiliary information.
\item \textbf{SR-IEM} \cite{pan2020rethinking} utilizes a modified self-attention mechanism to estimate the item importance and recommends based on the global preference and current interest.
\item \textbf{SR-GNN} \cite{wu2019session} as the first proposed GNN-based model, adopts gated GNNs to obtain item embeddings and recommends by generating the session representation with an attention mechanism.
\item \textbf{NISER+} \cite{gupta2019niser} extends SR-GNN by introducing L2 normalization, positional embedding, and dropout.
\item \textbf{SGNN-HN} \cite{pan2020star} extends SR-GNN by introducing a highway gate after the GNNs and a star node, which is connected with all items in the given session graph.
\end{itemize}

\subsubsection{Evaluation Metric}
Following \cite{wu2019session, pan2020star}, we use P@K (Precision) and MRR@K (Mean Reciprocal Rank) to evaluate the recommendation performance where K is 20.

\subsubsection{Parameter Setup}
We use the recent 10 items of the given session. We adopt an Adam optimizer with an initial learning rate $1e^{-3}$, $\beta_1 \!=\! 0.9$, $\beta_2 \!=\! 0.999$, and a decay factor 0.1 for every 3 epochs. L2 regularization is set to $1e^{-5}$. The batch size $N$ is set to 100 and the dimension of item embedding $d$ is set to 256. The weight of contrastive learning loss $\lambda$ is set to 0.7. The temperature parameters $\tau$ are set to 0.085 and 0.005 for main loss and contrastive loss. All ratios $\gamma$ for augmentation methods are set to 0.5. The factor for frequency of node connections $k$ is set to 0.75. All parameters are initialized using a uniform distribution with a range of $\left[ \frac{-1}{\sqrt{d}}, \frac{1}{\sqrt{d}} \right]$. In addition, the setup of each encoder (e.g., SR-GNN and NISER+) follows that of the corresponding paper.

\begin{table}[]
\centering
\resizebox{0.48\textwidth}{!}{%
\begin{tabular}{@{}cccccccc@{}}
\toprule
\multicolumn{2}{c}{\multirow{2}{*}{Method}}                                           & \multicolumn{2}{c}{Yoochoose1/64} & \multicolumn{2}{c}{Yoochoose1/4} & \multicolumn{2}{c}{Diginetica}  \\ \cmidrule(l){3-8} 
\multicolumn{2}{c}{}                                                                  & P@20            & MRR@20          & P@20            & MRR@20         & P@20           & MRR@20         \\ \midrule
\multirow{2}{*}{\begin{tabular}[c]{@{}c@{}}RNN-\\ based\end{tabular}} & GRU4REC       & 60.64           & 22.89           & 59.53           & 22.60          & 29.45          & 8.33           \\
                                                                      & CSRM          & 69.85           & 29.71           & 70.63           & 29.48          & 51.69          & 16.92          \\ \midrule
\begin{tabular}[c]{@{}c@{}}Attention-\\ based\end{tabular}            & SR-IEM        & 71.15           & 31.71           & 71.67           & 31.82          & 52.35          & 17.64          \\ \midrule
\multirow{3}{*}{\begin{tabular}[c]{@{}c@{}}GNN-\\ based\end{tabular}} & SR-GNN        & 70.57           & 30.94           & 71.36           & 31.89          & 50.73          & 17.59          \\
                                                                      & NISER+        & 71.27           & 31.61           & 71.80           & 31.80          & 53.39          & 18.72          \\
                                                                      & SGNN-HN       & \underline{72.06} & \textbf{32.61}  & \underline{72.85} & \underline{32.55} & \underline{55.67} & \underline{19.45}          \\ \midrule
\multirow{3}{*}{\begin{tabular}[c]{@{}c@{}}CL-\\ based\end{tabular}}  & SR-GNN w/ GCL & 71.16           & 31.27           & 71.79           & 31.42          & 52.73          & 17.90          \\
                                                                      & NISER+ w/ GCL & 71.64           & 32.08           & 72.30           & 32.00          & 54.74          & 19.26          \\
                                                                      & SR-GCL        & \textbf{72.14}  & \underline{32.33}           & \textbf{73.11}  & \textbf{32.70} & \textbf{55.93} & \textbf{19.53} \\ \bottomrule
\end{tabular}
}
\caption{Overall performance on three datasets. For each column, the bold-faced number is the best score and the second performer is underlined.}
\label{tb:overall_performance}
\end{table}

\subsection{Results and Discussions} \label{sec:res_dis}
In this section, we present the evaluation results. Except for the overall comparison result, we present the results only on Yoochoose1/64 and Diginetica.

\subsubsection{Overall Performance Comparison} \label{subsec:res_dis_rq1}
The overall experimental results are summarized in Table \ref{tb:overall_performance}. Although CSRM shows a great improvement from GRU4Rec in RNN-based models, SR-IEM outperforms those because its attention mechanism helps avoid the bias caused by unrelated items. SR-GNN shows a comparable result with SR-IEM by modeling the transition relationship of items. SGNN-HN, a state-of-the-art, shows significant improvements on all metrics by proposing a virtual node that connects items without direct connections. SR-GCL on top of the SGNN-HN encoder outperforms the state-of-the-art baselines except for MRR@20 on Yoochoose1/64. We also experimented by applying GCL to other encoders, and the results show better performance than the vanilla encoders. The lower result of MRR@20 on Yoochoose1/64 is repeatedly reported in the following experiments, so we will discuss this in detail at the end.

\begin{figure}[t]
    \centering
    \includegraphics[width=0.48\textwidth,keepaspectratio]{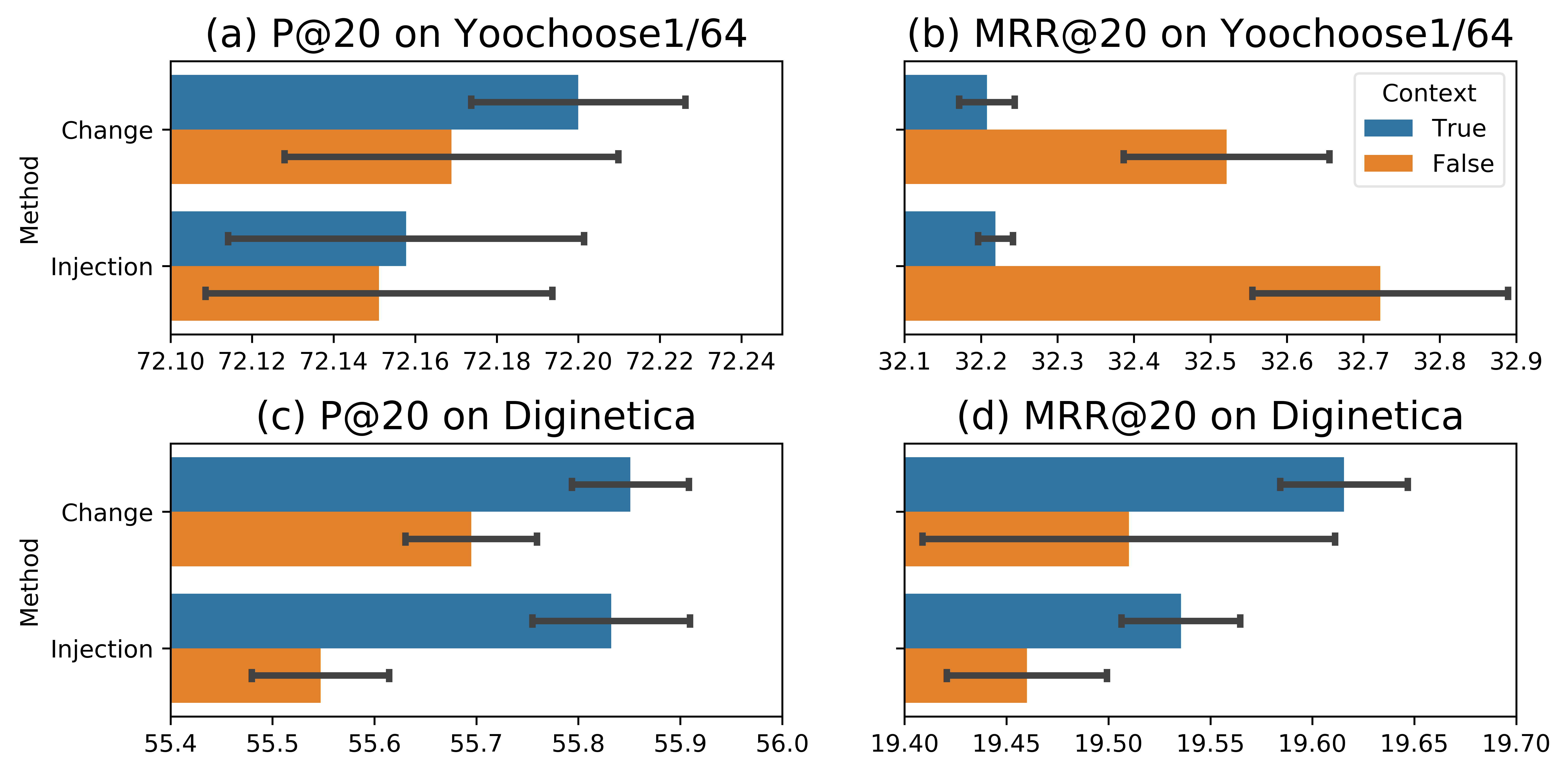}
    \caption{Comparison of considering the global context (True) and not (False).}
    \label{fig:random_noise}
\end{figure}

\subsubsection{Impact of Considering the Global Context} \label{subsec:res_dis_rq3}
To analyze the impact of considering the global context in our proposed augmentation approaches, we experimented with two extreme ways to find synonyms. The one is selecting from 1-hop neighbor nodes in the global item graph, which is currently adopted. The other one is selecting from all items regardless of their connections. In Figure \ref{fig:random_noise}, the bars are the means of nine performances when varying $\gamma$ is from 0.1 to 0.9. Each cap means the standard deviation of each case. The results have a similar tendency except Figure \ref{fig:random_noise} (b). The methods considering the global context (i.e., blue bars) show better performance than those which randomly select synonyms without the context (i.e., red bars). In addition, our proposed approaches have lower standard deviations as shown in the caps. Note that the way without the context is a similar operation to masking some items.

\begin{figure}[t]
    \centering
    \includegraphics[width=0.48\textwidth,keepaspectratio]{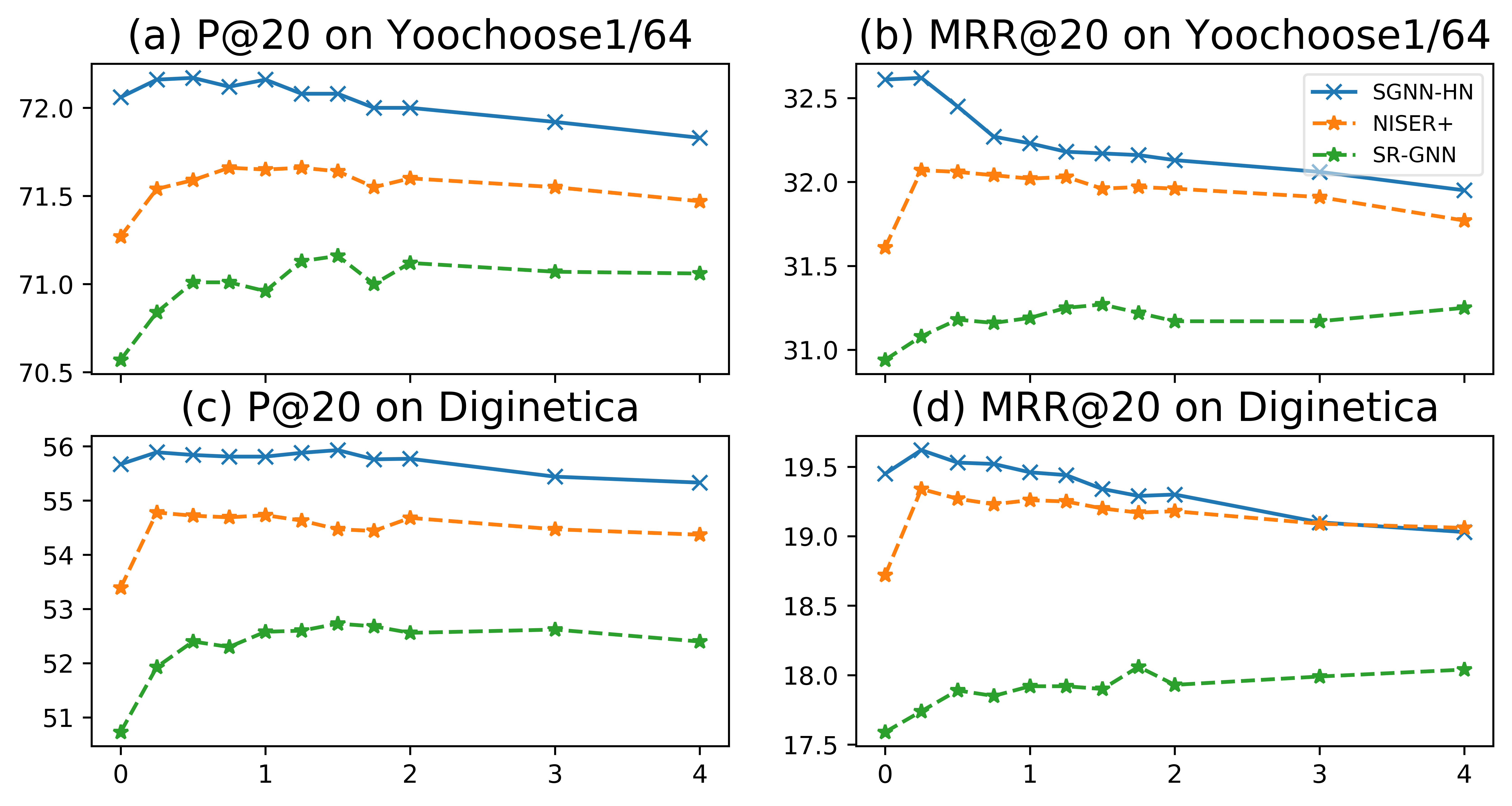}
    \caption{Performance varying a weight of contrastive loss $\lambda$ for three encoders (i.e., SGNN-HN, NISER+, and SR-GNN).}
    \label{fig:ratio_of_contrastive_learning}
\end{figure}

\subsubsection{Impact of Contrastive Loss}
We investigate the interaction effect between contrastive loss and prediction loss by varying the weight $\lambda$. We expect that contrastive learning as an auxiliary task is helpful for the recommendation performance, but it could degrade the performance if the effect is too large. So we experimented to find the optimal $\lambda$ for our task. In addition, we applied this experiment to other GNN-based encoders, SR-GNN and NISER+ to confirm that our framework is model-agnostic used for general session-based recommendations. Figure \ref{fig:ratio_of_contrastive_learning} shows that our proposed framework using SGNN-HN as an encoder achieves the best performance at less than 1. Then, as $\lambda$ increases in more than 1, the performance decreases. The case of using NISER+ has a similar result, but that looks relatively less sensitive to $\lambda$. When we replace the encoder with SR-GNN, the performance improvement is most prominent. There is no phenomenon that the performance is lowered as $\lambda$ increases.

\subsubsection{Discussions}
The experiment results show the superiority of SR-GCL in all evaluations except MRR@20 on Yoochoose1/64 consistently. We speculate the reason why SR-GCL is not effective in that case in two aspects. The first factor is the characteristics of the dataset. The ratio of a sequence length of 3 or less in Yoochoose1/64 is 63\%. Compared with that in Diginetica (i.e., 47\%), session augmentations are more difficult to work on Yoochoose1/64. In addition, the ratio for the number of neighbor items with 2 or less in Yoochoose1/64 is 31\%, indicating significantly lower connectivity compared with others. This may cause over-fittings, leading to degrading the performance when considering the global context (See Appendix \ref{Appendix_datasets_characteristics} for detailed statistics). The second one is the relationship between contrastive learning and metrics. In terms of P@20, contrastive learning introduces to explore a wider range of target items. However, this could conflict with the main task which adopts cross-entropy for one target, therefore, degrade MRR@20.

\section{Conclusion}
We present SR-GCL, a contrastive learning framework for an SBR to better capture the item and session representations, and propose two global context enhanced augmentation methods. For future work, we plan to devise better augmentation techniques suitable for shorter session data. Beyond the current contrastive learning framework where positive and negative pairs should be obtained, we plan to develop a non-contrastive learning method for an SBR where negative pairs are not required.

\section{Acknowledgments}
We thank Prof. Kyunghyun Cho (New York University) for a fruitful discussion and long-term collaboration.

\clearpage
\bibliography{Bibliography/aaai22}

\clearpage
\appendix
\setcounter{secnumdepth}{2}
\section*{Supplementary Materials}
\subsubsection{Organization}
The appendix is organized as follows. In Sections \ref{Appendix_SGNN}-\ref{Appendix_projection_head} for supplementing Section 'Method' of the main text, we describe the GNN-based encoder consisting of three procedures in detail, three conventional augmentations for session data, and Projection Head for contrastive learning. Then, for supplementing Section 'Experiments and Analysis', we further report the three experimental results and analysis in Section \ref{Appendix_experiments}. Finally, in Section \ref{Appendix_datasets_characteristics}, we present the data statistics in terms of distribution to complement our discussion of the main text.

\section{Graph Neural Network Based Encoder} \label{Appendix_SGNN}
There have been many models to learn a session representation for a recommendation \cite{hidasi2015session, wang2019collaborative, pan2020rethinking, wu2019session, gupta2019niser, pan2020star}. Among them, we leverage a GNN-based encoder which outperforms RNN-based models and shows decent improvements. In particular, we adopt Star Graph Neural Networks with Highway Networks (SGNN-HN) \cite{pan2020star} as a state-of-the-art model. We detail three processes, constructing a session graph, updating the graph, and obtaining the representation of the session as follows.

\subsection{Constructing a Session Graph}
We construct a local session graph $\mathcal{G} \!=\! \left( \mathcal{V}, \mathcal{E} \right)$ from each session $s$ to pass the graph-based model. $\mathcal{V}\!=\! \left\{ \left\{ v_1, v_2, ..., v_{\left\vert \mathcal{V} \right\vert - 1} \right\}, v_s \right\}$ includes all unique items in the session and a star node $v_s$. The $i$-th node embedding after passing the $l$-th layer of GNN is represented by a hidden state $h_i^l \in \mathbb{R}^d$. We initialize the hidden states as $h_i^0 \!=\! v_i$ and initialization for the hidden state of a star node $v_s$ is $h_s^0$, which is the average value of all $h_i^0$ as 
\begin{equation}
    h_s^0 = \frac{1}{N} \sum_{i=1}^N h_i^0
\end{equation}
where $N \!=\! \left\vert \mathcal{V} \right\vert - 1$ is the number of all unique items. A star node is connected to all nodes with bidirectional edges. Information from non-adjacent nodes can be propagated by taking the star node as an intermediate node \cite{pan2020star}. $\mathcal{E}$ is an edge set consisting of two types of edges. The first one is for connections of all nodes, which are converted into two normalized adjacency matrices (i.e., incoming matrix $A^I$ and outgoing matrix $A^O$) to pass the GNN. The second one is for connections of a star node.

\subsection{Updating a Graph}
First, nodes in a graph $\mathcal{G}$ are updated by propagating information of the neighbor nodes. $m_i^l$ denotes message (i.e., information from the neighbors) for a hidden state $h_i$ at the $l$-th layer, which can be formulated as
\begin{equation}
\begin{aligned}
    m_i^l = \left[ A_i^I(\left[ h_1^{l-1}, h_2^{l-1}, ..., h_N^{l-1} \right]^\top W^I + b^I)\; ; \quad \right. \\ 
    \left. A_i^O(\left[ h_1^{l-1}, h_2^{l-1}, ..., h_N^{l-1} \right]^\top W^O + b^O) \right]
\end{aligned}
\end{equation}
where N is the number of nodes, and $\left[ ; \right]$ denotes a concatenation. $A_i^I, A_i^O \in \mathbb{R}^{1 \times N}$ are the corresponding incoming and outgoing weights for the node $h_i$ (i.e., the $i$-th row of $A^I$ and $A^O$). $W^I, W^O \in \mathbb{R}^{d \times d}$ and $b^I, b^O \in \mathbb{R}^d$ are learnable parameters for the incoming and outgoing edges, respectively. So we feed the message $m_i^l$ and the previous state $h_i^{l-1}$ into the Gated Graph Neural Networks (GGNNs) \cite{li2015gated} for each node as
\begin{equation}
\begin{aligned}
    &z_i^l = \sigma(W_z m_i^l + U_z h_i^{l-1}),\\
    &r_i^l = \sigma(W_r m_i^l + U_r h_i^{l-1}),\\
    &\tilde{h}_i^l = tanh(W_h m_i^l + U_h (r_i^l \odot h_i^{l-1})),\\
    &\hat{h}_i^l = (1 - z_i^l) \odot h_i^{l-1} + z_i^l \odot \tilde{h}_i^l
\end{aligned}
\end{equation}
where $\sigma$ is a sigmoid and $\odot$ is an element-wise multiplication. $z_i^l$ and $r_i^l$ are an update gate and reset gate. $W_z, W_r, W_h \in \mathbb{R}^{d \times 2d}$ and $U_z, U_r, U_h \in \mathbb{R}^{d \times d}$ are trainable matrices. After the propagation of adjacent nodes, we also consider the overall information from the previous star node $h_s^{l-1}$. For each node, we decide how much information from the star node should be propagated with a gate network and attention mechanism as
\begin{equation}
\begin{aligned}
    &\alpha_i^l = \frac{(W_{q1}\hat{h}_i^l)^\top W_{k1}h_s^{l-1}}{\sqrt{d}}, \\
    &h_i^l = (1 - \alpha_i^l)\hat{h}_i^l + \alpha_i^l h_s^{l-1}
\end{aligned}
\end{equation}
where $W_{q1}, W_{k1} \in \mathbb{R}^{d \times d}$ are trainable matrices, and $\sqrt{d}$ is a scaling factor. After all hidden states are updated, the state of a star node is also updated. We apply an attention mechanism to compute different degrees of similarity for all node states $h^l = \left[ h_1^l, h_2^l, ..., h_N^l \right]$ by regarding the star node as a query. So the current state of the star node $h_s^l$ is computed as
\begin{equation}
\begin{aligned}
    &\beta^l = softmax \left(\frac{(W_{k2}h^l)^\top W_{q2}h_s^{l-1}}{\sqrt{d}}\right), \\
    &\qquad\qquad\qquad h_s^l = \beta^l h^l
\end{aligned}
\end{equation}
where $W_{q2}, W_{k2} \in \mathbb{R}^{d \times d}$ are learnable parameters, and $\beta^l \in \mathbb{R}^N$ are weights for all nodes. Multiple updating can be stacked to accurately represent the transition relationship between items. However, the more propagating information between nodes is repeated, the easier the model could lead to an over-fitting problem with the GNN \cite{qiu2019rethinking, pan2020star}. To address this problem, \cite{pan2020star} applied a highway gate for all nodes after the multiple updating. The highway gate computes the final hidden states $h^f$, which is the weighted sum of $h^0$ and $h^L$. $h^0$ and $h^L$ denote the hidden states before and after the $L$-layer GNN. The highway gate can be formulated as
\begin{equation}
\begin{aligned}
    & h^f = g \odot h^0 + (1 - g) \odot h^L, \\
    & \mbox{where } g = \sigma(W_g\left[ h^0 ;h^L \right])
\end{aligned}
\end{equation}
where $\sigma$ is a sigmoid function, $\left[ ; \right]$ denotes a concatenation, and $W_g \in \mathbb{R}^{d \times 2d}$ is a trainable matrix.

\subsection{Obtaining a Session Representation}
We obtain sequential item embeddings $u \in \mathbb{R}^{\left\vert s \right\vert \times d}$ from the corresponding nodes $h^f \in \mathbb{R}^{N \times d}$ which passed the GNN, where $\left\vert s \right\vert$ is the length of a session, and $N$ is the number of all unique items of a session. We add trainable position embeddings $p \in \mathbb{R}^{\left\vert s \right\vert \times d}$ to consider sequential information in an attention mechanism as
\begin{equation}
    u^p = u + p = \left[ u_1^p, u_2^p, ..., u_{\left\vert s \right\vert}^p \right]
\end{equation}
To represent the global preference of a session, a soft attention mechanism is applied with the current interest and a star node. This combines the items according to their degrees of preference as
\begin{equation}
\begin{aligned}
    &\gamma_i = W_0^\top \sigma(W_1 u_i^p + W_2 h_s^L + W_3 u_{\left\vert s \right\vert}^p + b_0), \\
    &\qquad\qquad\qquad \tilde{r} = \sum_{i=1}^{\left\vert s \right\vert} \gamma_i u_i^p
\end{aligned}
\end{equation}
where $W_0 \in \mathbb{R}^d$ and $W_1, W_2, W_3 \in \mathbb{R}^{d \times d}$ are learnable parameters, $b_0 \in \mathbb{R}^d$ is a bias, and $\sigma$ is a sigmoid function. That is, the global preference of the session $\tilde{r}$ is determined in consideration of the star node after the $L$-th layer $h_s^L$ and the last item $u_{\left\vert s \right\vert}^p$. We finally obtain the representation of the session $s$ to consider not only the global preference but also the current preference as
\begin{equation}
    r = \left[\tilde{r} ; u_{\left\vert s \right\vert}^p \right]
\end{equation}
where $\left[ ; \right]$ denotes a concatenation. $r \in \mathbb{R}^{2d}$ is used for a recommendation task and contrastive learning.

\section{Conventional Augmentations for a Session} \label{Appendix_augmentations}
For comparison, we also present three conventional methods including Crop, Mask, and Reorder \cite{xie2020contrastive}. Compared with our proposed methods, they could have little diversity of variation. In addition, the rate of information loss is high when the methods are applied.

\subsection{Item Crop}
This is a technique commonly used in CV, which is to obtain a subset by randomly cropping a part of an image \cite{takahashi2019data}. Applying this method to sequence data is equivalent to obtaining a continuous sub-sequence of a sequence. This augmentation method $a_{cr}$ for a sequence $s$ is formulated as
\begin{equation}
    s^{a_{cr}} = a_{cr}(s) = \left[ x_i, x_{i+1}, ..., x_{i+N_{cr}-1} \right]
\end{equation}
where $i$ means a starting index randomly selected, and $N_{cr} = \lfloor \gamma_{cr} \times \left\vert s \right\vert \rfloor$ is the length of the sub-sequence. It is effective to obtain a local view of the historical session.

\subsection{Item Mask}
This technique applies zero-masking to some parts of a sample. It has been widely adopted to avoid over-fittings in other fields, which is called word-dropout in NLP \cite{bowman2015generating} and node-dropout in GNN \cite{wu2021self, you2020graph}. To apply this technique for a session $s$, we randomly make a set $\mathcal{I}_{ma} = \left\{ i_1, i_2, ..., i_{N_{ma}} \right\}$ indicating masked indices, where $N_{ma} = \lceil \gamma_{ma} \times \left\vert s \right\vert \rceil$ is the number of masked items. The items pointed by $\mathcal{I}_{ma}$ are replaced with a special item [mask]. This method $a_{ma}$ can be formulated as
\begin{equation}
\begin{aligned}
    s^{a_{ma}} = a_{ma}(s) = \left[ \tilde{x}_1, \tilde{x}_2, ..., \tilde{x}_{ \left\vert s \right\vert } \right], \\
    \mbox{where } \tilde{x}_j = 
    \begin{cases}
    x_j \qquad\qquad \mbox{if } j \notin \mathcal{I}_{ma}\\
    \mbox{[mask]} \qquad\; \mbox{if } j \in \mathcal{I}_{ma}
    \end{cases}
\end{aligned}
\end{equation}
As items within a session implicitly represent the intention of a user, even if $s^{a_{ma}}$ has only several elements of the session, it could still reserve the main intention.

\subsection{Item Reorder}
This aims to shuffle a part of a given sequence. For sessions in the real-world, the order of users' interactions is not strictly enforced, but flexible due to various unobservable external factors \cite{tang2018personalized, covington2016deep}. To apply this method to a session $s$, we randomly shuffle a continuous sub-sequence $\left[ x_i, x_{i+1}, ..., x_{i+N_{re}-1} \right]$, where $i$ is a starting index randomly selected, and $N_{re} = \lceil \gamma_{re} \times \left\vert s \right\vert \rceil$ is the length of the sub-sequence. This augmentation method $a_{re}$ is formulated as
\begin{equation}
    s^{a_{re}} = a_{re}(s) = \left[ x_1, ..., \tilde{x}_i, ..., \tilde{x}_{i+N_{re}-1}, ..., x_{\left\vert s \right\vert} \right]
\end{equation}
where $\left[ \tilde{x}_i, \tilde{x}_{i+1}, ..., \tilde{x}_{i+N_{re}-1} \right]$ is a shuffled sub-sequence. With this method, we encourage our model to rely less on the order of a session. This enhances the model to be more robust when it encounters unexpected sessions.

\section{Projection Head} \label{Appendix_projection_head}
Projection Head aims to distinguish contrastive learning from the main task (i.e., next item prediction) by mapping representations to the latent space where contrastive loss is applied. Without this layer, contrastive learning could damage useful information for the main task because it is trained to be invariant to data transformation \cite{chen2020simple}. Therefore, we use another representation vector $z \in \mathbb{R}^d$ in the projected space through this module to separate from the main objective.

\section{Supplementary Experiments} \label{Appendix_experiments}

\begin{figure}[t]
    \centering
    \includegraphics[width=0.48\textwidth,keepaspectratio]{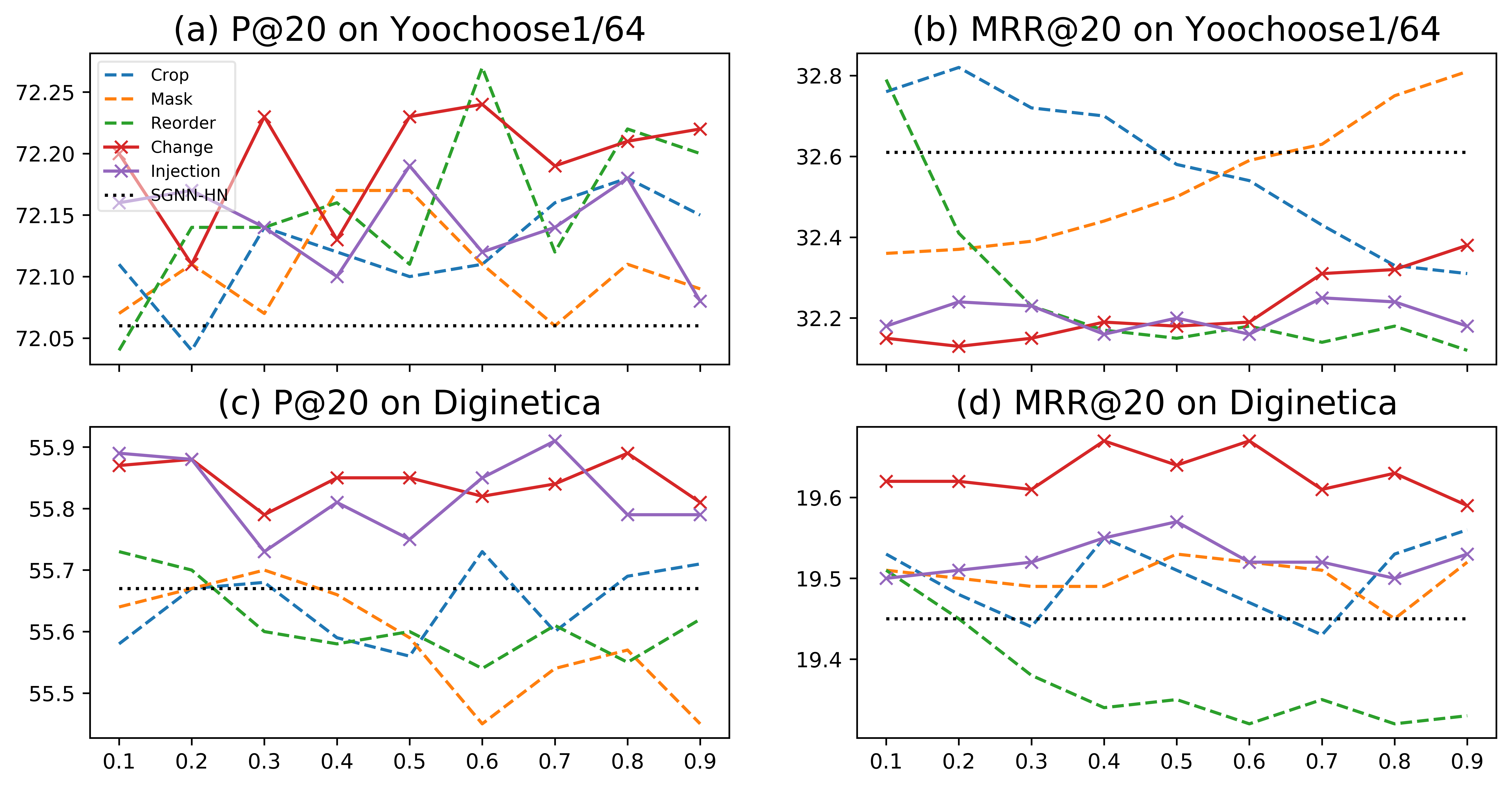}
    \caption{Performance varying an augmented ratio $\gamma$.}
    \label{fig:ratio_of_augmentations}
\end{figure}

\subsection{Impact of Individual Augmentation}
Figure \ref{fig:ratio_of_augmentations} shows the effect of individual augmentation operators by varying the ratio $\gamma$ from 0.1 to 0.9, where $\gamma$ means the degree of transformation of each session. If this value is large, it is difficult to have the same underlying semantics as the original session. Otherwise, if it is too small, the contrastive effect between differently augmented views of the same sample would be reduced. For the experiment, we use two augmented sessions generated by using the same method with a fixed ratio. Note that even if the same augmentation method is used, two generated sessions may be different due to the randomness of applied positions.

Taken overall, we can see that Change and Injection augmentations have better performance than other methods except for the case of Figure \ref{fig:ratio_of_augmentations} (b). Also, our proposed methods are relatively less influenced by $\gamma$. The reason is that even if the degrees of deformation are severe, they do not generate very different views from the original sample due to the consideration of the global context. In contrast, Crop and Mask are sensitive to $\gamma$. They also show the opposite tendency as $\gamma$ increases because they play a similar role in hiding information of a session. Meanwhile, Reorder shows a significant difference between P@20 and MRR@20. It maintains relatively low performance on MRR@20. The reason is that Reorder makes the current interest difficult to be known. To be specific, our model considers the last clicked-item as a proxy of current interest, but this method could shuffle the order of the sub-session including the last item, which dilutes the meaning.
Of course, other augmentations can also stochastically affect this, but Reorder can perturb these sequential dependencies more.
In summary, a high ratio of reordering could cause a mismatch between the global preference and the current interest.

\begin{figure}[t]
    \centering
    \includegraphics[width=0.48\textwidth,keepaspectratio]{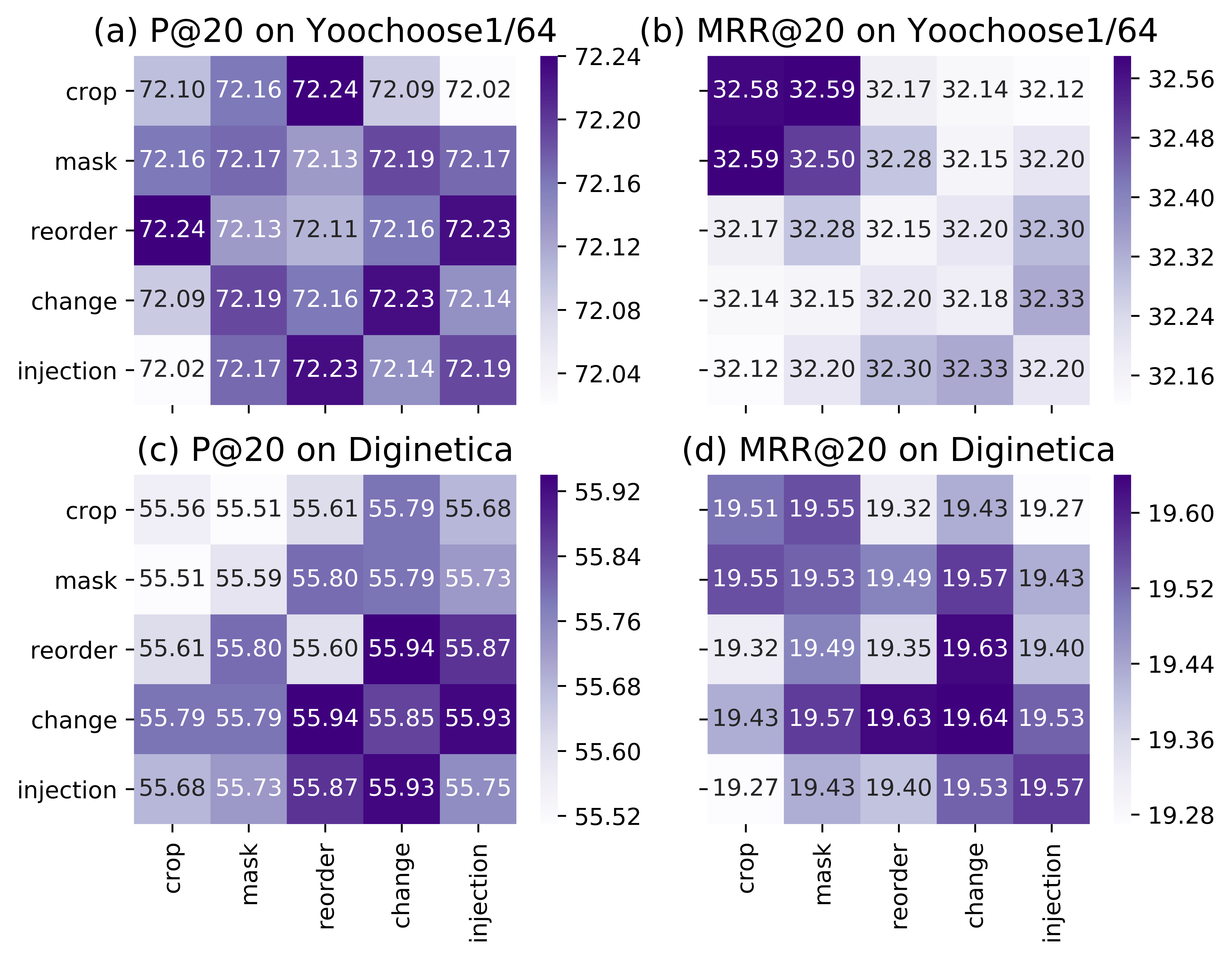}
    \caption{Heat maps for the combinations of two augmentation methods.}
    \label{fig:heatmap}
\end{figure}

\subsection{Combination of Augmentations} \label{subsec:res_dis_rq2}
Figure \ref{fig:heatmap} shows the performance according to the combinations of two augmentation methods. We use two augmented sessions generated by two fixed methods. The results show that exactly one combination cannot be the best for session data. However, our proposed methods (i.e., Change and Injection) can provide more options for session augmentations, which show the improved performance than the combinations of conventional methods (i.e., Crop, Mask, and Reorder) except for the case of Figure \ref{fig:heatmap} (b). It is notable that the combination of weak augmented methods (i.e., Change and Injection) and a strong method (i.e., Reorder) has better performance as mentioned in the advanced experiment.

\begin{figure}[t]
    \centering
    \includegraphics[width=0.48\textwidth,keepaspectratio]{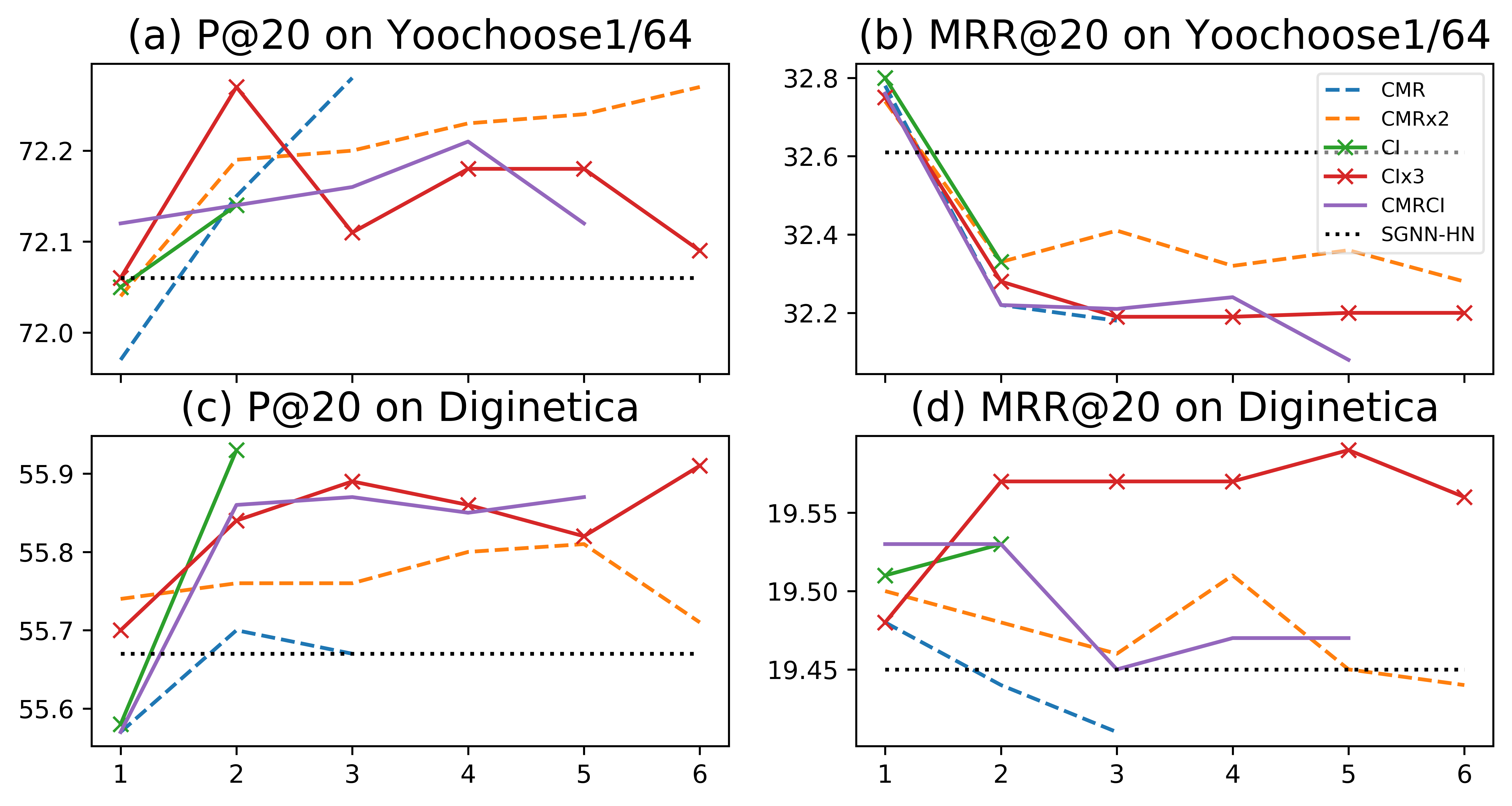}
    \caption{Performance varying the number of selected augmentation methods. CMR stands for Crop, Mask, and Reorder while Change and Injection are abbreviated to CI.}
    \label{fig:number_of_augmentations}
\end{figure}

\subsection{Impact of the Number of Selected Methods}
Our framework allows setting the number of selected methods flexibly due to our multi-positive loss described in the main text. This means that the number of negative samples of contrastive loss can be adjusted independently without changing the batch size. In this experiment, we tested how different the number of randomly selected methods impact the performance.

Figure \ref{fig:number_of_augmentations} shows the performance varying the number of selected methods $M$. When the batch size is $N$, the number of augmented representations used for contrastive learning per batch is $N \times M$. CMR means that we use an augmentation set including Crop, Mask, and Reorder. CI stands for Change and Injection. The cases except for Figure \ref{fig:number_of_augmentations} (b) show two results. First, the performance using the augmentation sets including CI is generally better than only CMR. Second, the number of selected methods does not need to be fixed to 2. Even though there is no optimal number of augmentations to ensure the best performance for all conditions, it can be optimized according to a specific purpose. The results could confirm that our contrastive learning framework is effective when the selected number is two or more, which means that there must be other views of the same session to take advantage of contrastive learning.

\begin{center}
\begin{table}[]
\resizebox{0.48\textwidth}{!}{%
\begin{tabular}{ c c c c }
\toprule
                        & Yoochoose1/64  & Yoochoose1/4  & Diginetica \\ \midrule
\# of clicks            & 557,248        & 8,326,407     & 982,961    \\
\# of train sessions    & 369,859        & 5,917,745     & 719,470    \\
\# of test sessions     & 55,898         & 55,898        & 60,858     \\
\# of items             & 17,745         & 30,470        & 43,097     \\
Avg. of session length  & 6.16           & 5.71          & 5.13       \\
\bottomrule
\end{tabular}%
}
\caption{Statistics of datasets.} \label{tb:data_stat}
\end{table}
\end{center}

\begin{figure}[t]
    \centering
    \includegraphics[width=0.48\textwidth,keepaspectratio]{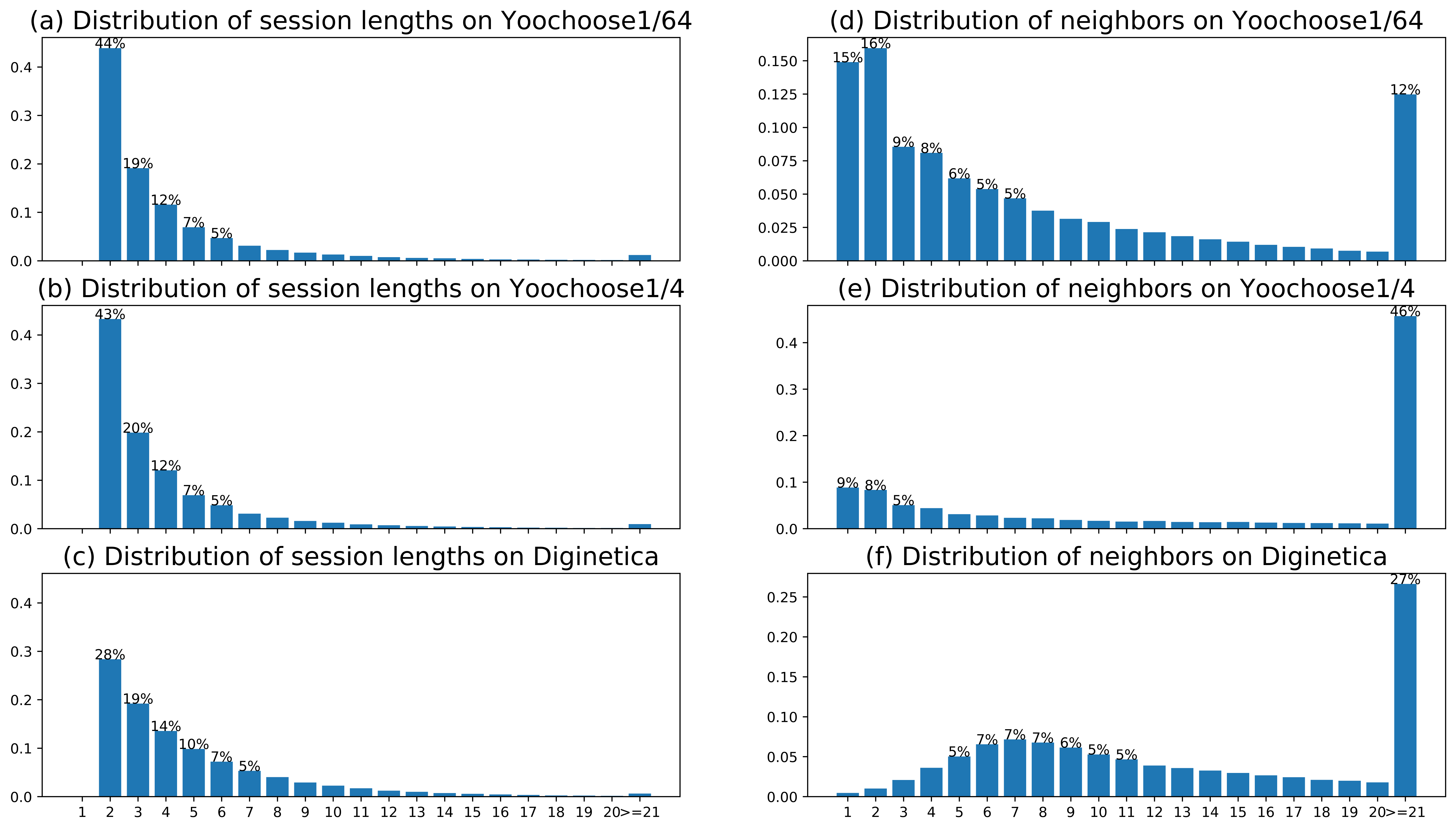}
    \caption{Distributions of session lengths and the number of neighbors on three datasets.}
    \label{fig:histogram_graph}
\end{figure}

\section{Inherent Characteristics of Datasets} \label{Appendix_datasets_characteristics}
In this section, we investigate the characteristics of the data itself from two aspects in order to analyze the tendency of the performance results. One is the distribution of session lengths, and the other is the distribution of the number of neighbors for each node.

Figure \ref{fig:histogram_graph} (a)-(c) show the distributions of session lengths on three datasets. Compared with the distribution on Diginetica, we can see that the ratio of short sessions in Yoochoose for both 1/64 and 1/4 is high. Specifically, the ratio of a session length of 3 or less in Yoochoose1/64 is 63\% while that in Diginetica is 47\%. By such data characteristics, when the session is short, it is difficult to apply augmentations. This is because no matter how much deformation there is, there is little room to be applied. This means that relatively insignificant contrastive effects could be obtained from the augmentations on Yoochoose1/64.

Meanwhile, the three datasets show different aspects regarding the number of neighbors on each item as shown in Figure \ref{fig:histogram_graph} (d)-(f). Yoochoose1/64 has significantly lower connectivity compared with others, meaning many isolated nodes. So it is easy for those many isolated nodes to find neighbors, which could be more vulnerable to over-fittings because the global context to be considered is limited. On the other hand, the neighbors of an item on Yoochoose1/4 and Diginetica are significantly distributed with higher values than those on Yoochoose1/64. This diversity of neighbors could help a model to be robust if we utilize the global context.

\end{document}